\documentclass[review]{elsarticle}

\usepackage[english]{babel}
\usepackage{listings} 
\usepackage[usenames]{color} 
\usepackage{lineno,hyperref}
\usepackage{url}
\usepackage{booktabs}
\usepackage{chngcntr}
\counterwithin{table}{section}
\modulolinenumbers[5]

\journal{Journal of \LaTeX\ Templates}
\newcommand{\var}{$s $ }
\newcommand{\varLabel}{$shrinking$ }
\newcommand{\varLbl}{$shrinking$}

\begin{document}

\begin{frontmatter}

\title{Time Aware Knowledge Extraction for Microblog Summarization on Twitter }

\author{Carmen De Maio}
\address[address1]{Deparment of Computer Science, University of Salerno, Fisciano (SA), Italy}
\ead{cdemaio@unisa.it}

\author{Giuseppe Fenza}
\ead{gfenza@unisa.it}

\author{Vincenzo Loia\corref{mycorrespondingauthor}}
\cortext[mycorrespondingauthor]{Corresponding author}
\ead{loia@unisa.it}

\author{Mimmo Parente}
\ead{parente@unisa.it}




\begin{abstract}
Microblogging services like Twitter and Facebook collect millions of user generated content every moment about trending news, occurring events, and so on. Nevertheless, it is really a nightmare to find information of interest through the huge amount of available posts that are often noise and redundant. In general, social media analytics services have caught increasing attention from both side research and industry. Specifically, the dynamic context of microblogging requires to manage not only meaning of information but also the evolution of knowledge over the timeline.
This work defines Time Aware Knowledge Extraction (briefly TAKE) methodology that relies on temporal extension of Fuzzy Formal Concept Analysis. 
In particular, a microblog summarization algorithm has been defined filtering the concepts organized by TAKE in a time-dependent hierarchy. The algorithm addresses topic-based summarization on Twitter. Besides considering the timing of the concepts, another distinguish feature of the proposed microblog summarization framework is the possibility to have more or less detailed summary, according to the user's needs, with good levels of quality and completeness as highlighted in the experimental results.
\end{abstract}


\end{frontmatter}


\section{Introduction}

\paragraph{Context} 
Nowadays, microblogging streams are useful to detect and track political events\cite{politicaleventdetection1}, media events\cite{mediaeventdetection1}, and other real world events\cite{eventdetection3}. 
Nevertheless, it is really difficult to understand the main aspects of the news or events inquiring these microblogging services. In fact, given a specific topic on Twitter a huge amount of relevant tweets that are redundant or not relevant due to the ambiguity and noise of the social media exists. Furthermore, the dynamic context of microblogging requires to manage not only meaning of information but also the evolution of knowledge over the time. To face with this side effect many applications have been realized on Twitter, like Tweetchup (tweetchup.com), Twitalyzer (twitalyzer.com), which provide social media analytics services to detect and track trending topics. Moreover, automatic microblog summarization algorithms that extend Latent Diriclet Allocation (LDA) exist that consider both chronological order of the tweets and their information content, but at two distinct stages \cite{hp, Sharifi, transaction}. 
In the light of the described scenario, this work defines topic-based microblog summarization framework extending Fuzzy Formal Concept Analysis to manage time relations among tweets and introducing two main distinguishing features, that are specifically: first considering both time and meaning of the tweet at the same time to analyze knowledge evolution over the timeline; and second providing summaries with different level of detail according to the user's needs exploiting the peculiar properties of timed fuzzy lattice.

\paragraph{Problem} Formally, this work tries to face with the following problem. Given a topic-focused timestamped tweet stream 
and a level of \varLabel \var,
the task is aimed to filter and chronologically order tweets in order to produce a Microblog Summary $MS_s$ that provides a complete description of the story covering main concepts describing topic development over the timeline. The proposed framework is able to retrieve more or less detailed summary according to the user's demand in terms of the closure level of \varLabel.

\paragraph{Proposed Solution} This work defines a Time Aware Knowledge Extraction (briefly, TAKE) as a new methodology to solve the problem of topic-based microblog summarization focusing here on Twitter to give experimental evidence. 
More specifically, the summarization process is achieved taking into account both semantics and timestamps of the tweets. Firstly, content will be annotated via sentence wikification that is the practice of representing a sentence with a set of Wikipedia concepts (i.e., entries)\cite{wikify1,wikify2}. 
Secondly, it is possible to identify temporal peaks of tweet frequency analyzing timestamps and exploiting the Offline Peak-Finding Algorithm (OPAD), proposed in \cite{twitinfo}. Then, taking into account the meaning of the tweet content and time dependences among detected peaks, temporal extension of Fuzzy Formal Concept Analysis \cite{ffca,temporal1} will be performed in order to arrange tweets into a hierarchy of time dependent concepts, that is a \emph{timed fuzzy lattice}. Finally, a summarization algorithm has been defined exploring resulting timed fuzzy lattice knowledge structure. The algorithm extracts chronologically ordered tweets summarizing main concepts of the story according to their temporal evolution.

\paragraph{Experimental Results} The proposed framework has been performed on the same tweet streams exploited in \cite{hp} that are focused on some real-world events, such as: Obamacare, Japan Earthquake, and so on. The results have been evaluated considering the following metrics: \emph{Novelty Measurements}, \emph{Text-based Coverage of Wikipedia}, and \emph{Concept-based Coverage of Wikipedia}.  The evaluation has been performed by varying level of \varLabel \var in $[0-1]$. For all of the used metrics the system produces good performances. Specifically, the algorithm outperforms the results in \cite{hp} in terms of \emph{Novelty Measurement} and \emph{Text-based Coverage of Wikipedia}. Furthermore, evaluating \emph{Concept-based Coverage of Wikipedia} setting level of \varLabel with values $\sim 0.9$ (that is a verbose summary), the algorithm outperforms the results shown in \cite{hp} in terms of F-Measure, with optimal Recall and comparable values of Precision.

\paragraph{Outlines} The manuscript is organized as follows: Section \ref{rw} provides an overview of the literature describing some related works; Section \ref{ffca} introduces the theoretical background, i.e. Fuzzy Formal Concept Analysis; Section \ref{framework} introduces the overall framework detailing each phase in the sections \ref{phase1}, \ref{take} and \ref{phase3}; finally, Section \ref{evaluation} shows the obtained results and argues the comparison with other existing approaches. 
\section{Related Works}
\label{rw}
Nowadays, automatic microblog summarization has caught increasing attention from worldwide researchers. 

From the time-dependent document summarization point of view, some existing approaches are aimed to address update summarization task defined in TAC (www.nist.gov/tac). Specifically, they emphasize the novelty of the subsequent summary \cite{li2006improving}. Unlikely, the proposed approach focuses more on the temporal development of the story (i.e. topic or event) that is stressed by the multitude of the messages posted through microblogging service, i.e. Twitter. 

From the microblog summarization point of view, some pioneering approaches working on Twitter exist that are essentially aimed to describe topic extracting list of relevant words or sentences. Specifically, TweetMotif \cite{twitmotif} summarizes what's happening on Twitter providing a list of relevant terms that should explain Twitter topics. \cite{sharifi2010automatic} and \cite{Sharifi} extract a succinct summary for each topic using a phrase reinforcement ranking approach. \cite{liu2011sxsw} explores tweets and linked web contents to discover relevant information about topics. Moreover, \cite{chakrabarti2011event} generates summaries especially for sport topics. Furthermore, \cite{Sharifi2} defines frequency and graph based method to select multiple tweets that conveyed information about a given topic without being redundant. Other approaches are based on integer linear programming \cite{Liu} or clustering to perform the summarization of Evolving Tweet Streams \cite{Shou}. 
Other approaches consist of aggregating tweets about specific topic into a visual summaries. These visualizations must be interpreted by users and do not include sentence-level textual summaries. For instance, Visual Backchannel \cite{backchannel} and TwitInfo \cite{twitinfo} allow users to graphically browse a large collection of tweets. Specifically, \cite{backchannel} visualizes conversations in Twitter data using topic streams that is visually represented as stacked graphs and  TwitInfo \cite{twitinfo} uses a timeline-based display that highlights peaks of high tweet activity.

Considering our proposal we find some similarities in \cite{hp} and in \cite{transaction}. 
Specifically, \cite{hp} describes a framework for summarizing events from tweet stream. The authors define two topic models, Decay Topic Model (DTM) and Gaussian DTM, to extract summaries from microblog, and they finally argue that these models outperforms LDA (Latent Dirichlet Allocation) baseline that doesn't consider temporal relation among tweets. 
Instead, the approach used in \cite{transaction} introduces a sequential summarization for Twitter trending topics exploiting two  approaches: a stream based approach that is aimed to extract important subtopic concerning with specific category (e.g., News, Sport, etc.) identifying peak areas according to the timestamps of the tweets; and a semantic based approach leveraging on Dynamic Topic Modeling, that extends LDA in order to consider timeline, to identify topic from a semantic prospective in the time interval. In \cite{transaction} the authors argue that hybrid approach that considers stream and semantic of the tweets outperforms other ones.

In general, these research works highlight that to achieve microblog summarization, due to the dynamic nature of its content,  it is crucial to consider both the chronological order of the posts and their information content. Unlike these microblog summarization approaches that consider the time and meaning of the tweets at two different stages, our solution considers both timestamps and meaning of the tweets at the same time. This work presents the Time Aware Knowledge Extraction (briefly TAKE) methodology, as a new approach to perform conceptual and temporal data analysis of tweets' content for microblog summarization. TAKE extends Fuzzy Formal Concept Analysis \cite{ffca} introducing time dependencies among objects, in order to provide a summary that follow the evolution of the story over the timeline. Furthermore, the proposed framework reveals good performances in terms of F-Measure, with optimal Recall and comparable values of Precision with respect to the compared approaches. Specifically, the timed fuzzy lattice extracted by TAKE enable us to support user requests providing less or more succinct summary according to the specific needs.

\section{Theoretical Background: Fuzzy Formal Concept Analysis}\label {ffca}
The formal model behind the proposed methodology for microblog summarization is the fuzzy extension of Formal Concept Analysis (briefly, Fuzzy FCA or FFCA) \cite{FCA}. FCA  is a theoretical framework which supplies a basis for conceptual data analysis, knowledge processing and extraction. Fuzzy FCA \cite{ffca} combines fuzzy logic into FCA representing the uncertainty through membership values in the range [0, 1]. 

Following, some definitions about Fuzzy FCA are given.

\emph{\bf Definition 1: }
{\it A {\bf  Fuzzy Formal Context} is a triple K =} ({\it G,M, I = $\varphi(G \times M))$, where  G  is  a  set  of  objects,  M  is  a  set  of  attributes,  and  I  is  a  fuzzy  set  on domain  G $\times$ M.  Each  relation  }({\it g, m}{\it )  $\in$  I  has  a  membership  value  $\mu(g, m)$  in [0, 1].}

\emph{\bf Definition 2: }	{\bf Fuzzy Representation of Object}. { \it Each object O in a fuzzy formal context K can be represented by a fuzzy set $\Phi(O)$ as $\Phi(O)$=\{A$_1(\mu_1)$, A$_2(\mu_2)$,\dots, A$_m(\mu_m)$\}, where \{A$_1$, A$_2$,\dots, A$_m$\} is the set of attributes in K and $\mu_i$ is the membership of O with attribute A$_i$ in K. $\Phi(O)$ is called the fuzzy representation of O.}

\begin{figure}[!h]
   \centering
  \includegraphics[width=\linewidth]{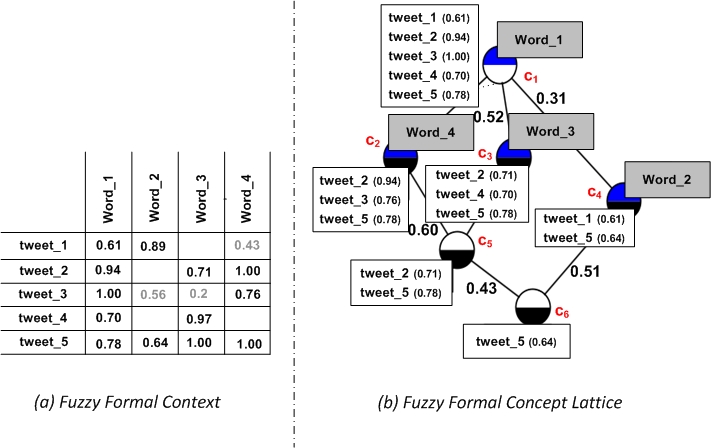}
  \caption{Portion of fuzzy formal context (a) and the relative concept lattice with threshold T = 0.6 (b)}
  \label{fig:ffca}
\end{figure}

Unlike FCA that use binary relation to represent formal context, Fuzzy Formal Context enables the representation of the fuzzy relation between objects and attributes in a given domain. So, fuzziness enables to model relation among object and attribute in a more smoothed way ensuring more precise representation and uncertainty management. Fuzzy Formal Context (see Definition 1) is often represented as a cross-table as shown in Figure \ref{fig:ffca}(a), where the rows represent the objects, while the columns, the attributes. Let us note that each cell of the table contains a membership value in [0, 1]. Specifically, Fuzzy Formal Context shown in Figure  \ref{fig:ffca}(a)  has a confidence threshold T=0.6, that means all the relationship with membership values less than 0.6 are not shown. 

Taking into account Fuzzy Formal Context, Fuzzy FCA algorithm is able to identify Fuzzy Formal Concepts and subsumption relations among them. More formally, the definition of Fuzzy Formal Concept and order relation among them are given as follows:

\emph{\bf Definition 3: }	{\bf Fuzzy Formal Concept}. {\it Given  a  fuzzy  formal  context  K=}({\it G, M, I}= $\varphi(G \times M))${\it  and  a  confidence threshold  T,  we  define  A$^{*}$= \{m $\in$  M $|$ $\forall$ g $\in$ A:  $\mu(g,  m)$ $\geq$  T \}  for  A $\subseteq$  G  and B$^{*}$= \{g $\in$  G $|$ $\forall$ m $\in$ B:  $\mu(g, m)$  $\geq$  T\}  for  B $\subseteq$  M.  A  fuzzy  formal  concept  (or fuzzy  concept)  of  a  fuzzy  formal  context  K  with  a  confidence  threshold T  is a pair} ({\it A$_f$ = $\varphi(A)$}{\it , B}){\it , where A  $\subseteq$ G, B $\subseteq$ M , A*=B  and B*=A. Each object g $\in$   $\varphi(A)$ has a membership $\mu_g$ defined as}

\centerline{$\mu_g$= min $_{m \in B}   \mu(g, m)$ }
        
\noindent {\it where $\mu(g, m)$  is the membership value between object  g  and attribute m, which is defined in I. Note that if B=\{ \} then $\mu_g$ = 1 for every g. A and B are the extent and intent of the formal concept }({\it $\varphi(A),B)$ respectively}.

\emph{\bf Definition 4: }	{ \it	Let} ({\it A$_1$, B$_1$}){\it and} ({\it A$_2$, B$_2$}) {\it be two fuzzy concepts of a fuzzy formal context} ({\it G, M, I} ).  ($\varphi$({\it A$_1$}), {\it B$_1$}){\it is  the  \textbf{subconcept}  of } ({\it $\varphi$}({\it A$_2$}){\it ,B$_2$}){\it ,  denoted as  }({\it $\varphi$}({\it A$_1$}){\it , B$_1$}) $\leq$ ($\varphi$({\it A$_2$}),{\it B$_2$}){\it ,  if  and  only  if  $\varphi$}({\it A$_1$}) $\subseteq$  $\varphi$({\it A$_2$}) ($\Leftrightarrow$ {\it B$_2$ $\subseteq$  B$_1$}){\it . Equivalently, }({\it A$_2$, B$_2$}){\it \  is the \textbf{superconcept} of }({\it A$_1$, B$_1$}).

Let us note that each node (i.e. a formal concept) is composed by the objects and the associated set of attributes, emphasizing by means fuzzy membership the object that are better represented by a set of attributes. In the figure, each node can be colored in different way, according to its characteristics: a half-blue colored node represents a concept with $own$ attributes; a half-black colored node instead, outlines the presence of $own$ objects in the concept; finally, a half-white colored node can represent a concept with no $own$ objects (if the white colored portion is the half below of the circle) or attributes (if the white half is up on the circle). 

An example of Fuzzy Formal Concept is $C_4$  that is composed of objects $A_f = {tweet_1, tweet_5}$ and attributes $B = {``word_1, word_2'', \dots}$) with $\mu_{tweet_1}$= 0.61  and  $\mu_{tweet_5}$= 0.64 , as shown in \ref{fig:ffca}(b). Furthermore, Fuzzy FCA carries out  Fuzzy Concept Lattice, i.e. a hierarchycal structure of the concepts according to the order relation (see Definition 4), as shown in Figure \ref{fig:ffca}(b). For instance, let us observe in Figure \ref{fig:ffca}(b), the concept $c_5$ is \textit{subconcept} of the concepts $c_2$ and $c_3$. Equivalently the concepts $c_2$ and $c_3$ are \textit{superconcepts} of the concept $c_5$. 

Now, it is possible to define Fuzzy Concept Lattice as follows:

\emph{\bf Definition 5: }	{\it 	\textbf{A Fuzzy Concept Lattice} of a fuzzy formal context K with a confidence threshold T  is a set F}({\it K}){\it of all fuzzy concepts of K with the partial order $\leq$ with the confidence threshold T }.

Figure \ref{fig:ffca}(b) shows an example of lattice coming from the related table, with threshold $T = 0.6$. In fact,  FCA provides also an alternative graphical representation of tabular data that is somewhat natural to navigate and use \cite{FCA}. Furthermore, the Fuzzy FCA introduces the definition of  Fuzzy Formal Concept Similarity. 

\emph{\bf Definition 6: }	{\it 	\textbf{ Fuzzy Formal Concept Similarity} between concept $K_1 = (\mu(A_1), B_1)$ and its subconcept $K_2 = (\mu(A_2), B_2)$ is defined as

$E(K_1, K_2) = \frac{\left | \varphi \left ( A_1 \right )\bigcap \varphi \left ( A_2 \right )\right | }{\left | \varphi \left ( A_1 \right )\bigcup \varphi \left ( A_2 \right )\right |}$ }

where $\bigcap$ and $\bigcup$ refer intersection and union operators\footnote{The fuzzy intersection and union are calculated using \textit t-norm and \textit t-conorm, respectively. The most commonly adopted \textit t-norm is the minimum, while the most common \textit t-conorm is the maximum. That is, given two fuzzy sets \textit A and \textit B with membership functions $\mu_A(x) and \mu_B(x), \mu_{A\bigcap B}(x) = min(\mu_A(x),\mu_B(x)) and \mu_{A\bigcup B}(x) = max(\mu_A(x),\mu_B(x))$.} on fuzzy sets, respectively.

On one hand, the FCA provides a taxonomic arrangement of concepts and extracts the subsumption relationships (often known as a ``hyponym-hypernym or is-a relationship") among them. On the other hand, Fuzzy FCA enables to considers these relations with a certain degree of truth (i.e., an approximate subsumption). In other words, the resulting fuzzy lattice elicits data-driven knowledge-based, hierarchical dependences, refining the taxonomic nature of this structure weighting interrelation among concepts introducing Fuzzy Formal Concept Similarity as stated in Definition 6.

\section{Framework Overview}
\label{framework}
The proposed framework is aimed to address microblog summarization service on twitter. Specifically, this work defines a novel Time Aware Knowledge Extraction (briefly TAKE) methodology aimed to perform temporal and conceptual data analysis to foster dynamic nature of social media introducing intelligent analytics services. In particular we show how tweet stream will be analyzed to extract meaning of the tweets and to detect temporal correlation among them. 

\begin{figure}[t!]
   \centering
  \includegraphics[width=\linewidth]{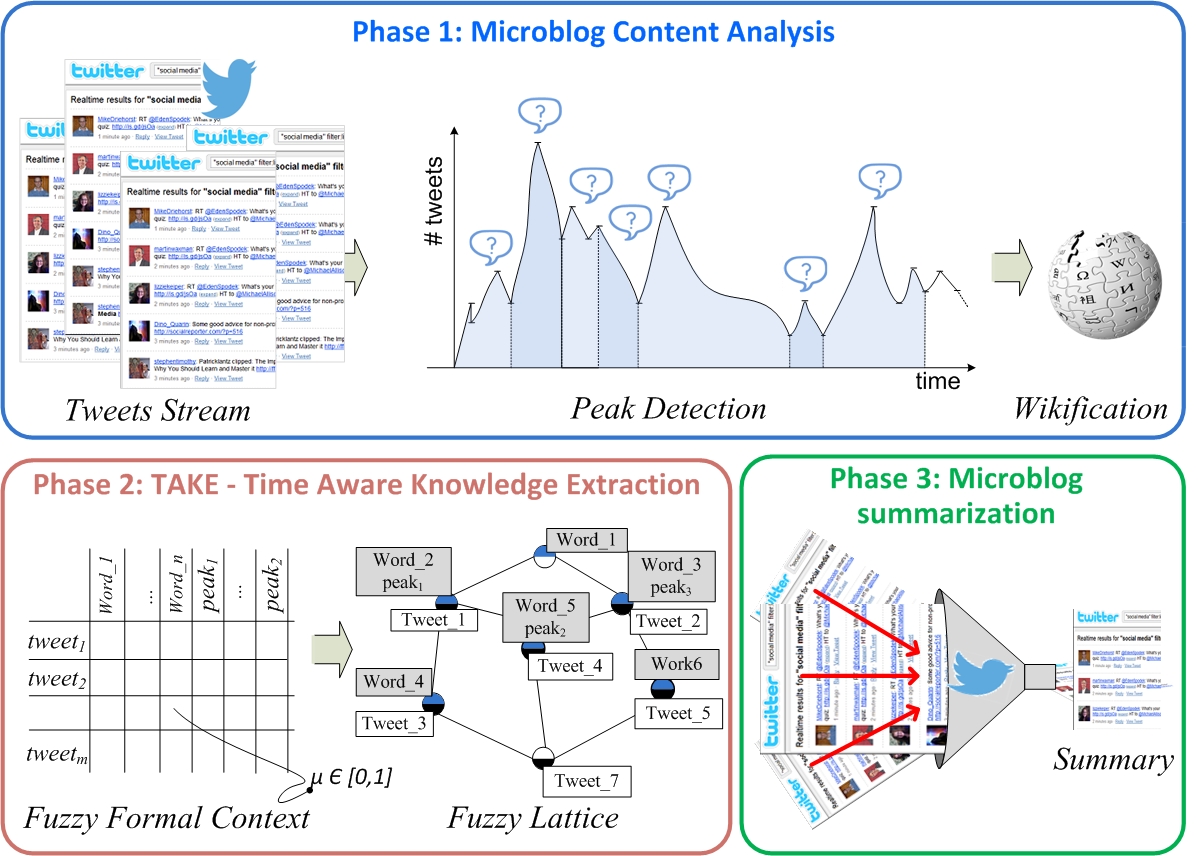}
  \caption{Overall Process of the framework}
  \label{fig:all}
\end{figure}

Specifically, Figure \ref {fig:all} sketches the whole process of the system that is composed of following main phases: 
\begin{itemize}
\item [-] \textit{Microblog Content Analysis} (see Section \ref{phase1}). It takes as input a tweet stream and detects tweet frequency peaks, then performs tweet's features extraction exploiting text analysis services, such as wikification,
determining the meaning of the tweet and performing ad-hoc term weighting;
\item [-] \textit{TAKE - Time Aware Knowledge Extraction} (see Section \ref{take}). It takes as input term weighted tweets and their timestamps and performs Time Aware FFCA in order to arrange tweets into a hierarchy carrying out also time dependence relation among extracted concepts;
\item [-] \textit{Microblog Summarization Algorithm} (see Section \ref{phase3}). It is a summarization algorithm that given the timed fuzzy lattice resulting by TAKE extracts a filtered set of tweets that covers the key concepts of the story considering the timeline and the \varLabel level specified as input (See later for the definition and discussion of \varLabel).
\end{itemize} 

It is possible to distinguish phases performed online and offline. Specifically, \textit{Microblog Content Analysis} and \textit{Time Aware Knowledge Extraction} will be periodically performed offline also because they are time consuming activities. Instead, \textit{Microblog Summarization Algorithm} is performed at execution time according to the user request in terms of topic and level of \varLbl. Additional technical and formal details about each macro-phases are given in the next sections.

\section{Microblog Content Analysis}
\label{phase1}
This phase is aimed to characterize tweets extracting representing features considering both the timestamp and the meaning of the tweet. Specifically, this activity is preliminary to map the domain data (e.g., tweets content) into a fuzzy formal context, enabling FFCA execution (details are given in the following sections). 

This phase is composed of these steps:

\begin{itemize}
\item [-] \textit{Peak Detection}, to detect temporal peaks from Twitter streams;
\item [-] \textit{Content Wikification}, to identify and extract relevant features that characterize meaning of the input tweet;
\item [-] \textit{Inverse Tweet Frequency}, to measure how important a concept is.
\end{itemize} 

The mathematical modeling of the fuzzy formal context needs the representative features capable to represent both meaning and time dependencies among the tweets. The goal is to exploit a vector-based representation of each tweet and then build the matrix which represents the fuzzy formal context. This matrix will show the relationships (in terms of degree values) between the extracted features (i.e., peaks and wikipedia entities) and tweets in the application domain. Further details are given in the following subsections.

\subsection{Peak Detection}\label{peak}
This step identifies temporal peaks in tweet frequency exploiting the \textit{Offline Peak-Finding Algorithm} (OPAD) (listing 1), proposed in \cite{twitinfo}. The algorithm is based on the idea of TCP congestion control, which uses a weighted moving  mean and variance to determine if there is a new peak area \cite{twitinfo}. 

Given a time-sorted collection of tweets, the algorithm locates surges by tracing tweet volume changes. Let $T=(t_1, t_2, ..., t_n)$ a time-sorted collection of tweets, we group tweets that are posted within the same 1440 minute (i.e., 1 day) time window. At this point we have a list of tweet counts $C=(C_1, C_2,... C_t)$ where $C_i$ is the number of tweets in bin $i$. The objective is identify each bin $i$ such that $C_i$ is large relative to the recent history $C_{i-1}, C_{i-2}, \dots C_{1}$.

\noindent Initializing the mean and variance with the first time interval (line 2-3), the algorithm loops through the whole tweet stream (line 5). If the number of tweets in the current bin (i.e., $C_i$) is greater than $\tau$ (we use $\tau = 2$) mean deviations from the current mean (i.e., {$\frac{\left | C_i - mean \right |}{meandev}>\tau$}), and the tweet number in current bin is increasing (i.e., $C_i>C_{i-1}$, line 6), then a new peak window starts (line 7). Then, the algorithm will loop until the condition $C_i>C_{i-1}$ is verified and updates the mean and variance (line 8-11). So, the peak search stops when the tweet number in the bin is less than the number of the previous one. After that, in the loop of lines 12-20 the bottom of peak interval is searched, which occurs either when the tweet number in the current bin is smaller than the tweet number at starting of the peak window (line 12) or another significant increase is found (line 13). At line 23, new peak window is included in the set of found peak areas. Every time we iterate over a new bin count, we update the mean and mean deviation (lines 9, 17 and 24) by means of \textit{Update} function (line 30-34). In the function \textit{Update} $\alpha$ is set to 0.125 as in \cite{twitinfo}. 

\lstset{
language=JAVA,
    basicstyle=\scriptsize,
    numbers=left,
    numberstyle=\tiny,
    stepnumber=1,
    numbersep=0pt,
    frameround=ftff,
    frame=lines,
    belowcaptionskip=.25\baselineskip,
}
\lstset{
literate={C1}{{$C_1$}}{1}
{Cp}{{$C_p$}}{1} 
{Ci}{{$C_i$}}{1}
{C2}{{$C_{i-1}$}}{1}
{Cs}{{$C_{start}$}}{1}
{aa}{{$\alpha$}}{1}
{frac1}{{$\frac{\left | C_i - mean \right |}{meandev}>\tau$}}{1}
}
\begin{lstlisting}[caption={OPAD- Offline Peak Area Detection}]
  windows = []
  mean = C1
  meandev = variance ( C1, ..., Cp )

  for i = 2; i < len(C); i++ do
     if      frac1      and Ci >  C2  then
        start = i-1
        while i < len(C) and Ci >  C2  do 
           (mean, meandev) = update(mean, meandev, Ci ) 
            i + +
        end while
        while i < len(C) and Ci >   Cs   do 
           if      frac1      and Ci >  C2  then
                end = - - i
                break
           else
                 (mean, meandev) = update(mean, meandev, Ci ) 
                 end = i + +
           end if 
        end while
        if ( Ci <   Cs  )  then 
           end = i - -
        windows.append(start, end)
     else
         (mean, meandev) = update(mean, meandev, Ci )
     end if
  end for 
  return windows

  function update(oldmean, oldmeandev, updatevalue):
     diff = |oldmean - updatevalue|
     newmeandev = aa*diff+(1-aa)*oldmeandev
     newmean = aa*updatevalue+(1-aa)*oldmean
  return (newmean, newmeandev)
\end{lstlisting}
Then the i-th tweet will be annotated temporally, such as follows:
\begin{itemize}
\item [-]  \textit{$tweet_i$ =  $\{\left \langle peak_i \right \rangle$}\}.
\end{itemize}

\subsection{Content Wikification} \label{wiki}
The previous step of Microblog's content Analysis process involves the extraction of concepts from an unstructured text in the tweet content. To achieve this aim this work exploits common-sense knowledge available in Wikipedia. In order to do this, the tweet content is wikified to extract a set of \textit{$\left \langle topic, relevance \right \rangle$} pairs corresponding to Wikipedia articles that are related to the tweet content itself with a specific relevance degree \cite{wikify1}.  
In particular, topics returned by applying the wikification upon a tweet content helped us to characterize the given text. 

Let us report an example by considering the following tweet:\\
\textit{$tweet_i$ =  ``President Obama just designated the largest marine reserve in the world"}.

\begin{figure}[b!]
   \centering
  \includegraphics[width=\linewidth]{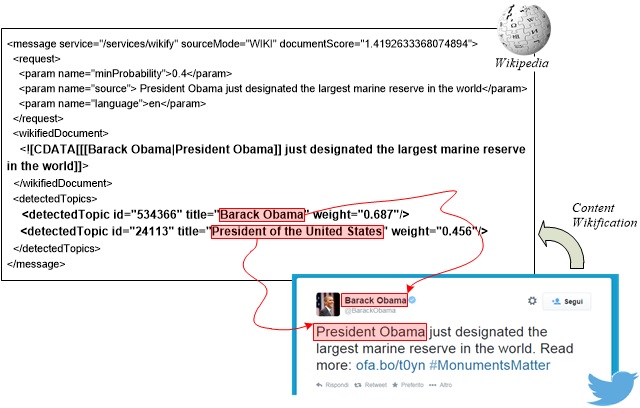}
  \caption{Example of tweet's content wikification.}
  \label{fig:wikify}
\end{figure}

The wikification process extracts from the above text a set of \textit{$\left \langle topic, relevance\right \rangle$} pairs. These pairs are features characterizing meaning of the input text. Taking into account the example above, the extracted topic (shown in Figure \ref{fig:wikify}) are:  
\centerline{$\left \langle Barack\ Obama, 0.678 \right \rangle$, $\left \langle President\ of\ the\ United\ States, 0.456 \right \rangle$}

%

Then, at this point, considering the example defined in Section \ref{peak} about $tweet_i$, the content will be annotated via sentence wikification as:

\centerline{\textit{$tweet_i$ = $\{\left \langle peak_i \right \rangle\} \bigcup $}} 
\centerline{\textit{$\{\left \langle topic_{i_1}, relevance_{i_1} \right \rangle$, $\left \langle  topic_{i_2}, relevance_{i_2}\right \rangle, \dots, \left \langle topic_{i_m}, relevance_{i_m}\right \rangle\}$}}

\noindent where $m$ is the number of topics detected by sentence wikification of the $tweet_i$.

\subsection{Inverse Tweet Frequency}
After having analyzed the peak area which the tweets belong to (see Section \ref{peak}) and the wikification of tweet content (see Section \ref{wiki}), \textit{ITF (i.e., Inverse Tweet Frequency)} is exploited to refine membership of relevance degree of the topic found inside the tweet. It intuitively evaluates the measure of how much information each extracted topic (see Section \ref{wiki}) provides whether it is common or rare across all tweets. 
Specifically, let $ W= \{w_1, w_2, . . . , w_n\}$ be the set of topics extracted by means of wikification process from set of tweets $T=\{t_1, t_2, . . .,t_m\}$. Let us compute the ITF for each one topic as:\\
\centerline{$ \mathrm{itf}(w_i, T) =  \log \frac{N}{|\{t_j \in T: w_i \in t_j\}|}$}
where:
\begin{itemize}
\item [-]  N: total number of tweets analyzed;
\item [-] $|\{t_j \in T: w_i \in t_j\}| $: number of tweet from which the topic $w_i$ has been extracted.
\end{itemize}

This value is exploited to compute the final value that characterizes the frequency associated to each topic extracted for a tweet. In particular, the final relevance \textit{$f_{rel}$} associated to the topic $w_i$ with respect to the tweet $t_j$ is defined as:
\centerline{$f_{rel}(w_i, t_j) = relevance(w_i, t_j) \times itf(w_i, T);$}

Then, at this point, considering the example defined in previous sections about $tweet_i$, the content will be annotated as:\\
\centerline{\textit{$tweet_i$ = $\{\left \langle peak_i \right \rangle\} \bigcup  \{\left \langle topic_j, f_{rel_j} \right \rangle, \left \langle topic_{j+1}, f_{rel_{j+1}}\right \rangle, \dots, \left \langle topic_m, f_{rel_m}\right \rangle$}\}}

\section{TAKE - Time Aware Knowledge Extraction}
\label{take}
Time Aware Knowledge Extraction is an important feature to perform conceptual data analysis taking into account temporal relation among resources and to consequently carry out temporal correlation among concepts in order to represent their development over the timeline. The proposed approach to address this aim relies on  Fuzzy Formal Concept Analysis, but as stated in Section \ref{ffca}, FFCA does not cover time dependences in the data.

 In literature, some approaches that extend formal concept analysis to handle temporal properties and represent temporally evolving attributes exist \cite{TFCA}. Specifically, this temporal extension has been applied in \cite{TFCAchat} to search pedophiles on
the Internet analyzing chat conversation over the time. Here we adopt a distinct approach by extending FCA introducing fuzziness and temporal correlation among objects, in order to extract temporal dependencies among attributes in the concepts. 
  
This work defines a time extension of \textit {FFCA} to extract hierarchically and  temporal related concepts. Indeed, besides classical contexts, timed \textit {FFCA} extracts chronological relations among formal concepts inferred by analyzing time dependences among formal objects. 

From a theoretical viewpoint, this work extends FFCA to consider timeline defining special attributes for representing time relations among formal objects. Formally, a time aware fuzzy formal context is defined as follow:

\emph{\bf Definition 7: }	{\it A \textbf{ Time Aware Fuzzy Formal Context} is a fuzzy formal contexts $K_t$ = $(G, M\textsuperscript{+} = M \bigcup T, I_M= \varphi(G \times M), I_T)$, where $T$ is the set of time attributes and $I_T$ is a binary time relation $I_T \subseteq G \times T$ representing the relation between formal object $g \in G$ and time attributes $t \in T$. }

\begin{figure}[!h]
   \centering
  \includegraphics[width=\linewidth]{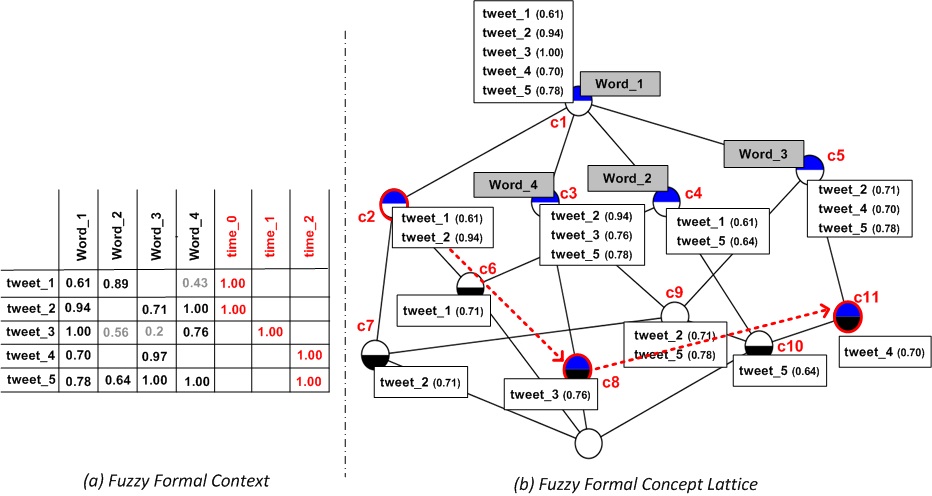}
  \caption{Time Aware Fuzzy FCA: portion of fuzzy temporal fuzzy formal context (a) and the relative temporal fuzzy concept lattice}
  \label{fig:tfca}
\end{figure}

For instance, if $g\in G$ and $t \in T$ are in relation $I_T$ means that $g$  happens at time $t \in T$. 

Time extension of Fuzzy FCA allows to organize tweets in a weighted hierarchical knowledge structure, that is a timed fuzzy lattice. In particular, a straight mapping defines a correspondence between the set of attributes M and linguistic terms extracted from tweets content, as well as the set G of objects and the tweets collection. 

Let us consider timed fuzzy formal context and correspondent timed fuzzy lattice in Figure \ref{fig:tfca}. Specifically, Figure \ref{fig:tfca}(b) emphasizes that each node (i.e., a formal concept) includes the objects, attributes and time attributes. For example in the lattice in Figure \ref{fig:tfca}(b), a concept is ($A_f= {tweet_1, tweet_2}$, $B = {``word_1, time_0''}$) with $\mu_{tweet_1}$= 0.61  and  $\mu_{tweet_2}$= 0.94.

The resulting timed fuzzy lattice emphasizes a temporal correlation among concepts and highlights how the concepts change over the timeline (Figure \ref{fig:tfca}). To represent the concept development over the timeline in a timed fuzzy lattice have been introduced temporal edges (in Figure \ref{fig:tfca} red dashed arrows) among related concepts.  The temporal edges allow the evolution of attributes to be followed over the time. A temporal precedence relation is defined over time points. The direction of the arrow indicates this precedence. In the lattice in Figure \ref{fig:tfca}(b), the evolution of attributes  is represented as: $c_2 \rightarrow c_8 \rightarrow c_{11}$, i.e., $\{Obama\} \rightarrow \{Obama, election\} \rightarrow \{Obama, President\}$.
  

\section{Microblog Summarization Algorithm}
\label{phase3}
The microblog summarization algorithm has been defined walking across concepts of the timed fuzzy lattice structure resulting from Time Aware Knowledge Extraction. The general idea behind is to explore fuzzy formal concepts according to the chronological order of the peak areas. The algorithm incrementally selects the \emph{best tweet}, that is the tweet with highest degree of membership belonging to the most representative concept $C$, at each exploration stage. The most representative concept is one that has highest weight $w(C)$ defined later. 

More formally, given  a  fuzzy  formal  context  K=({\it G, M, I}= $\varphi(G \times M))$ and
fuzzy formal concept {\it C = ($\varphi(A)$}{\it,B)} (see Section \ref{ffca}), the level of \varLabel of formal concept $C$ is defined as: 

\centerline{$s(C) = 1 - support(C) = 1 - \frac{|A|}{|G|}$}

\noindent where $|A|$ is the number of tweets in $C$ and $|G|$ is the number of tweets in the overall stream.
\noindent Then, the weight $w(C)$ of fuzzy formal concept $C$ will be evaluated as follows:

\centerline{$w(c)=\frac{\sum_{m\in B}\mu_m }{|B|}$}

\noindent where $|B|$ is the number of attributes in $C$ and the membership $\mu_m$ is defined as follows:

\centerline{$\mu_m=max_{g \in \varphi(A)}\mu(g,m)$}

\noindent where $\mu(g, m)$  is the membership value between object  $g$ and attribute $m$ (see Section \ref{ffca}).
%
%
%
%
%
%
%

The microblog summarization algorithm is detailed in the Listing \ref{listSum}. First of all, the sets of covered attributes (i.e., $\it CA$ ), covered concepts (i.e., $\it CC$ ) and summary (i.e., $\it MS_s$ ) are initialized as empty set (line 4-6). Then, the algorithm selects concepts of the timed fuzzy lattice whose \varLabel is greater than a threshold $s$ specified as input (line 8). After that, the algorithm sorts peak areas in a descending order, that is the most recent peak area will appear first (line 9). Finally, the algorithm loops across each concepts that have been grouped by peak area $p_i \in P$ (line 10). At each itearion, the algorithm selects the most representative concept $c_{max}$ (line 14) and the \emph{best tweet} $t_{max}$ with highest degree of membership belonging to $c_{max}$ (line 15). At the end of each iteration, the algorithm includes $t_{max}$ in the resulting summary (i.e., $MS_s$) (line 16) and updates the set of covered attributes $CA$ (line 17) and the set of covered concepts $CC$ (line 18). 
\lstset{
language=JAVA,
    basicstyle=\scriptsize,
    numbers=left,
    numberstyle=\tiny,
    stepnumber=1,
    numbersep=0pt,
    breaklines=true,
    frameround=ftff,
    frame=lines,
    belowcaptionskip=.25\baselineskip,
}

\lstset{
literate=
{_p}{{$_p$}}{1} 
{_1}{{$_1$}}{1}
{_n}{{$_n$}}{1} 
{_i}{{$_i$}}{1}
{_j}{{$_j$}}{1}
{_2}{{$_2$}}{1}
{_in}{{$\in$}}{1}
{_p1}{{$_{p1}$}}{1}
{_pi}{{$_{pi}$}}{1}
{_p2}{{$_{p2}$}}{1}
{_max}{{$_{max}$}}{1}
{_pn}{{$_{pn}$}}{1}
{_g}{{$\mu_g$}}{1}
{_cup}{{$\cup$}}{1}
{_attr}{{$\mu $m$  \in$ c$_{max}$}}{1}
{_oo}{{$\O$}}{1}
{_L}{{$\it L$}}{1}
{_S}{{$\it MS_s$}}{1}
{_P}{{$\it P$}}{1}
{_A}{{$\it A$}}{1}
{_B}{{$\it B$}}{1}
{_CA}{{$\it CA$}}{1}
{_CC}{{$\it CC$}}{1}
{_CP}{{$C^*_{p_i}$}}{1}
{_r}{{\var}}{1}
{_label}{{\varLabel}}{1}
{_supp}{{$\it s(c_i) = 1 - support (c_i)$}}{1}
{_forall}{{$\forall$}}{1}
{_sse}{{$\Leftrightarrow $}}{1}
{_*}{{$C^*$}}{1}
{_neq}{{$\neq $}}{1}
{_argmax}{{$argmax_{c_i \in (C^*_{p_i}\setminus CC)} \left (w(c_i)=\frac{\sum_{m\in (B\setminus CA)}\mu_m }{|B|}\right )$}}{1}{_argmaxt}{{$argmax_{g \in c_max}  \left (\mu_g \right )$}}{1}
}
\begin{lstlisting}[caption={Summarization by Time Aware FFCA}, label=listSum]
 Input: timed fuzzy lattice _L, peak areas _P, and    _label    level _r.
 Output: a microblog summary  _S  of _T tweets;

  _CA  = _oo, 
  _CC  = _oo,  
  _S  = _oo
 
  _* = {c_i _in _L |  	 _supp   	   > _r }
 _P  = (p_1, p_2,..., p_n) _forall i,j: p_i > p_j _sse i < j
  _CP = {c_i _in  _* | c_i = (_A,_B), p_i _in _B }
  
 for i = 1; i < len(_P); i++ do
   while  _CP \ _CC _neq _oo
      c _max  =                   _argmax
      t _max  =        _argmaxt
       _S  =  _S  _cup t _max
       _CA  =  _CA  _cup {m |    _attr    } 
       _CC  =  _CC  _cup c _max
    end while
  end for
 \end{lstlisting}


\begin{figure}[b!]
   \centering
  \includegraphics[width=\linewidth]{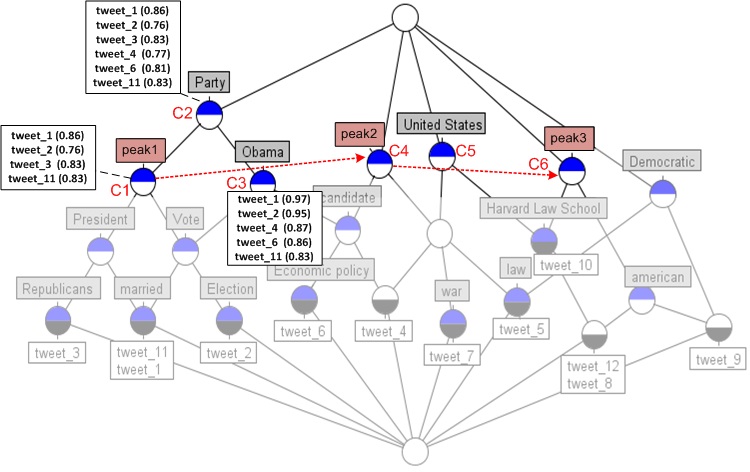}
  \caption{Example of timed Fuzzy Concept Lattice}
  \label{fig:lattice}
\end{figure}

Just to give an example, let us suppose an input level of \varLabel \var = 70\%. Figure \ref{fig:lattice} shows the concepts with level of \varLabel greater than 70\% resulting from the execution of line 8 in Listing \ref{listSum}. Table \ref{tabAlg} lists the set of candidate concepts grouped by peak area to which they belong to. 

\begin{table}[h]
\caption{Concepts of the timed fuzzy lattice in Figure \ref{fig:lattice} grouped by peak areas}\label{tabAlg}
\centering
\begin{tabular}{ll}
\hline
\textbf{peak}          & \textbf{Concepts} \\ \hline
\multicolumn{1}{l|}{1} & $c_1$, $c_2$, $c_3$        \\
\multicolumn{1}{l|}{2} & $c_2$, $c_3$, $c_4$, $c_5$   \\
\multicolumn{1}{l|}{3} & $c_5$, $c_6$           
\end{tabular}
\end{table}
According to the algorithm, the set of concepts is analyzed starting from the most recent peak area, that is $peak_1$. For each concept the weight $w$ is calculated and the concept with maximum value  of $w$ will be selected  (see Listing \ref{listSum}, line 14). Let us consider the following example: 

\centerline{$w(c_1) = 0.89; \ \ w(c_2)= 0.74; \ \ w(c_3)=0.94$;}

\noindent The first selected concept will be $c_3 = \{tweet\_1, tweet\_2, tweet\_11, tweet\_4,$ $ tweet\_6\}$ with maximum weight $w = 0.94$. The attributes covered by the concept $c_3$ are: \textit{Obama, Party}. 
 
The algorithm exploits fuzzy membership corresponding to the tweets in the selected concept (i.e., $c_3$) in order to look for the tweets with maximum membership degree. Thus, the tweet that will be introduced in the summary is $tweet\_1$ with highest degree of membership (i.e., $0.97$) belonging to the concept $c_3$ (see Listing \ref{listSum}, line 15). After updating summary including this tweet, the weight of remaining concepts will be updated removing attributes already covered by selecting $c_3$. In this case, the weight of remaining concepts is $0$ for both  $c_1$ and $c_2$. So, there are no more concepts to select in the peak area $peak 1$. So, the algorithm proceeds with next peak area, i.e. $peak_2$. At the end of the execution the resulting summary will be composed of following tweets:

\centerline  {$MS_s= \{tweet\_1, tweet\_7\}$.}

\section{Framework Evaluation}
\label{evaluation}
This section details the experimental results obtained performing the proposed summarization algorithm on specific tweet streams. As said before, the summarization algorithm relies on timed fuzzy lattice of tweets resulting from Time Aware Knowledge Extraction methodology execution. Since the timed fuzzy lattice allows to perform the summarization algorithm with different levels of \varLbl, the results have been evaluated by varying these levels. In particular, the higher the level of \varLabel the more detailed the resulting summary will be, that is the summary will include a greater amount of tweets. 

The discussion will continue as follows: description of the dataset of tweets (i.e. tweet streams) on which the framework has been executed (Section \ref{dataset}), definition of evaluation measures (Section \ref{measures}), and finally the experimental results will be discussed (Section \ref{results}).

\subsection{Tweet Streams}
\label{dataset}

The summarization framework has been applied on tweet streams focused on four real-world events\footnote{Specifically, the data have been provided by authors of \cite{hp}}: Facebook IPO\footnote{\url{http://en.wikipedia.org/wiki/Initial\_public\_offering\_of\_Facebook}}, Obamacare\footnote{\url{http://en.wikipedia.org/wiki/Patient\_Protection\_and\_Affordable\_Care\_Act}}, Japan Earthquake\footnote{\url{http://en.wikipedia.org/wiki/2011\_T\%C5\%8Dhoku\_earthquake\_and\_tsunami}} and BP Oil Spill\footnote{\url{http://en.wikipedia.org/wiki/Deepwater\_Horizon\_oil\_spill}}. 
The number of tweets for these events ranges from 9.570 tweets for Facebook IPO to 251.802 tweets for the Japan Earthquake. Specifically, Table \ref{tab} synthesizes how many tweets are included in each tweets stream. Let us note that the multitude of tweets related to each event highlights that nowadays microblogging summarization as well as other social media analytics services are welcomed to foster social media usage.

\begin{table}[h]
\label{tab}
\caption{Number of tweets for each dataset} 
\centering
\begin{tabular}{@{}ll@{}}
\toprule
\textbf{Name}         & \textbf{\# Tweets} \\ \midrule
\multicolumn{1}{l|}{Facebook IPO} & 9.570                    \\
\multicolumn{1}{l|}{Obamacare}&136.761                     \\
\multicolumn{1}{l|}{Japan Earthquake} &251.802                    \\
\multicolumn{1}{l|}{BP Oil Spill} & 79.676                    \\ \bottomrule
\end{tabular}
\end{table}

\subsection{Measurements}
\label{measures}
The proposed framework has been evaluated considering the following metrics:
\begin{itemize}
\item \emph{Novelty Measurements}, specifically \emph{Sequence Novelty Measurement} introduced in \cite{transaction} and \emph{Historical Novelty Measurement}.
\begin{itemize}
\item \emph{Sequence Novelty Measurement} measures average novelty among chronologically adjacent tweets included in the resulting summary. Information content \emph{I} has been used to measure the novelty of update summaries. In particular, it is defined as the average of \emph{I} increments of two adjacent new tweets added to summary. 
\begin{equation}
Novelty=\frac{1}{\left | D \right |-1}\sum_{i>1}\left ( I_{d_{i}}-I_{d_{i},d_{i-1}} \right )
\end{equation}
where:
\begin{itemize}
\item [-]$| D|$ is the number of the tweets in the generated summary;
\item [-]$I_{d_{i}}$ number of concepts in $d_{i}$;
\item [-]$I_{d_{i},d_{i-1}}$ cardinality of intersection of $d_{i},d_{i-1}$.
\end{itemize} 
\item\emph{Historical Novelty Measurement} evaluates average novelty among each tweet and all previous ones included in the resulting summary. This measure has been defined in this work to represent the update summary ratio considering history of chronologically previous tweets included in the generated summary. Analogously to the \emph{Sequence Novelty Measurement}, information content \emph{I} has been used to measure the novelty of update summaries. In particular, it is defined as
\begin{equation}
Novelty=\frac{1}{\left | D \right |-1}\sum_{i>1}\left ( I_{d_{i}}-\left [ \left ( \bigcup_{k<i}I_{d_{k}} \right )\bigcap I_{d_{i}} \right ]\right )
\end{equation}
where:
\begin{itemize}
\item [-]$| D|$ is the number of new tweets added in the summary;
\item [-]$I_{d_{i}}$ number of concepts in $d_{i}$;
\item [-]$I_{d_{i},d_{i-1}}$ cardinality of intersection of $d_{i},d_{i-1}$;
\item [-]$I_{d_{k}}$ with $k=1... |D|$ correspond to all tweets in the summary.
\end{itemize}
\end{itemize} 

\item \emph{Text-based Coverage of Wikipedia}, introduced in \cite{hp} where is called \textit{Quantitative Comparison with Wikipedia}, evaluates how much generated summary covers the gold one at text-level (i.e., considering n-grams). Specifically, gold summaries are extracted from Wikipedia\footnote{Indeed, gold summaries have been provided by authors of \cite{hp}. They are extracted considering the references of the relevant news articles cited in Wikipedia article corresponding to the topic/event of the tweet stream. For each of the Wikipedia references for the selected events, we extract the headline text which gives a one line summary of the corresponding news article.}.
Specifically, the metric counts the total number of \emph{n-grams} (excluding stop-words) in the generated summary $S^{gen}$ that are also included in the  gold summary $S^{gold}$. Let us define $NG_n^{gold}$ the set of n-grams in the gold summary and $NG_n^{gen}$ the set of n-grams in the generated summaries, this metric has been evaluated as follows:
\begin{equation}
g_n=\frac{1}{\left | NG_n^{gold} \right |} \sum_{ng\in NG_n^{gold}}min\left ( \left | ng\in NG_n^{gold} \right |,\left | ng \in NG_n^{gen} \right | \right )
\end{equation}

\begin{equation}
Sim\left ( S^{gold}, S^{gen} \right )=0,2\cdot g_1+0,3\cdot g_2+0,5\cdot g_3
\end{equation}
First equation calculates the number of n-grams common to both $S^{gold}$ and $S^{gen}$. In order to not let few frequent n-gram to dominate the counts, each n-gram is limited to the minimum number of counts between the  gold summary and the generated summary. The other equation calculates the final similarity score between the summaries by aggregating the number of matched 1, 2 and 3-grams. The weights allocated are meant to give a higher importance to 3-grams and lower importance to 1-grams. 

\item \emph{Concept-based Coverage of Wikipedia}, this metric has been defined in this work to evaluate how much the generated summary covers the gold summary at concept level. Indeed, each sentence of gold and generated summaries will be annotated via wikification that is the practice of representing a sentence with a set of Wikipedia concepts \cite{wikify1,wikify2}. More formally, let $C_{gold} =\{c_1, c_2, . . . , c_m\}$ and $C'_{gen} =\{c'_1,c'_2, . . . , c'_n\}$  be, respectively, the set of concept extracted from the sentences included in the gold summary and the set of concepts extracted from the generated summary. Then \emph{Concept-based Coverage of Wikipedia} will be evaluated in terms of well-known F-Measure that is obtained by combining measures of  Precision and Recall. Specifically, Precision and Recall will be evaluated as follows:
\begin{equation}
P=\frac{\left | C_{gold}\bigcap C'_{gen} \right |}{ \left | C'_{gen} \right |}  \ \ \ \ \ \ \ R=\frac{\left |  C_{gold} \bigcap  C'_{gen}  \right |}{ \left | C_{gold} \right |}
\end{equation}
Then, F-measure \emph{F} is computed as follows: 
\begin{equation}
F=2\times \frac{P\times R}{P+R}  
\end{equation}
So, this measure provides qualitative (i.e., Precision) and quantitative (i.e., Recall) information about how much generated summary covers the gold summary, and so, it evaluates semantically performances of the proposed microblog summarization approach. 
\end{itemize}

\subsection{Experimental Results}
\label{results}
The selected tweet streams and measures have been used to evaluate both the proposed approach (i.e., referred as TAKE) and methods defined in \cite{hp}. Since TAKE produces different summaries with different level of \varLbl, the results have been evaluated by varying the level of \var in $[0-1]$ and for all of the used metrics the system reveals good performances. 

\begin{figure}[b!]
\centering
\includegraphics[width=\textwidth]{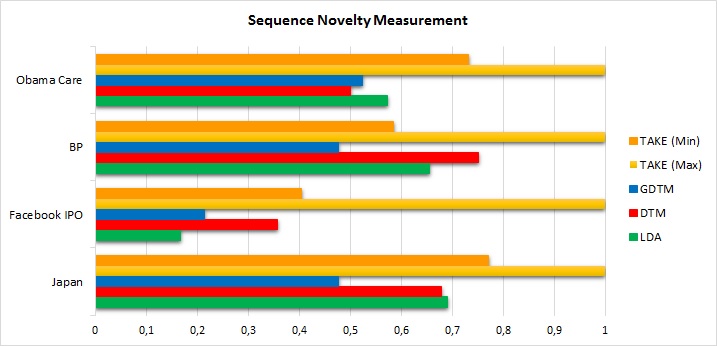}
\caption{Sequence Novelty Measurement Results.}\label{fig:seqNov}
\end{figure}

\begin{figure}[t!]
\centering
\includegraphics[width=\textwidth]{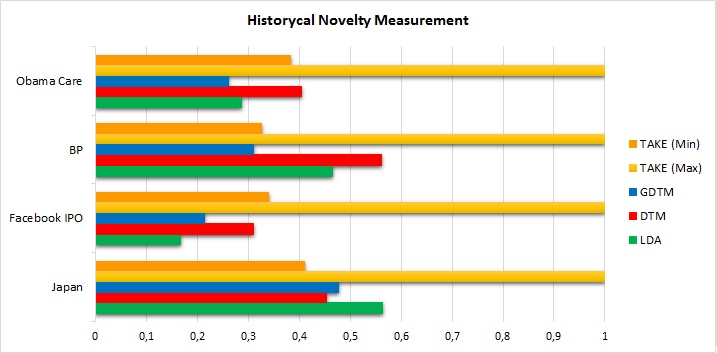}
\caption{Historical Novelty Measurement Results.}\label{fig:hisNov}
\end{figure}

\subsubsection{Novelty Results}
Figures \ref{fig:seqNov} and Figure \ref{fig:hisNov} show the results about novelty, respectively \emph{Sequence Novelty Measurement} and \emph{Historical Novelty Measurement}.
The results have been grouped by tweet streams (i.e., real world events of Facebook IPO, Japan Earthquake, and so on) and for each evaluated approach they are shown with different colors. In particular, TAKE has been evaluated by varying level of \varLabel \var in $[0-1]$ and plotting the obtained minimum and maximum values for both novelty measures.

Figure \ref{fig:seqNov} illustrates the results of novelty among adjacent tweets, that is \emph{Sequence Novelty Measurement}. On the one hand, it points out that the proposed approach produces summaries with maximum values of novelty highest than other approaches for each tweet stream. On the other hand, TAKE produces summaries with minimum values of novelty lower than other approaches only for the tweet stream of \emph{BP Oil Spill}. 

Analogously, the results of \emph{Historical Novelty Measurement} shown in Figure \ref{fig:hisNov} highlight that TAKE produces summaries with maximum values of novelty highest than other approaches for each tweet stream. Minimum values of \emph{Historical Novelty Measurement} produced by TAKE are enough close to the results obtained with other approaches, and so they are acceptable results. 

Since, the proposed microblog summarization returns chronological ordered tweets starting from the most recent ones, the results of \emph{Novelty Measurement} points out that TAKE generates more or less shorten summaries with acceptable levels of redundancy. Thus, the proposed method incrementally includes tweets in the resulting summary introducing significant amount of novel concepts improving the description of the event according to its development over the timeline.

\begin{figure}[b!]
\centering
\includegraphics[width=\textwidth]{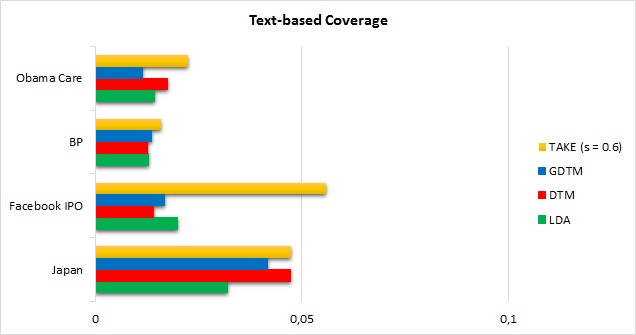}
\caption{\emph{Text-based Coverage of Wikipedia} Results.}\label{fig:text_coverage}
\end{figure}
\subsubsection{Text-based and Concept-based Coverage Results}
\begin{figure}[t!]
\centering
\includegraphics[width=\textwidth]{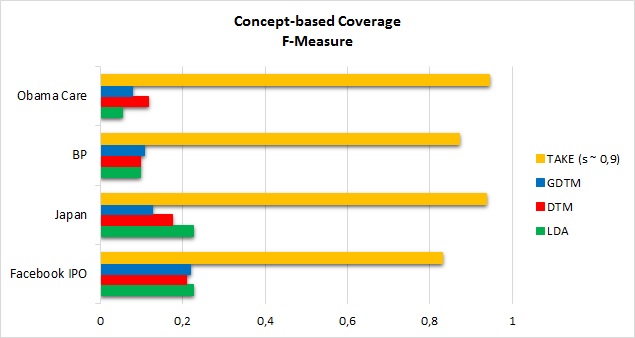}
\caption{Concept-based Coverage of Wikipedia Results.}\label{fig:concept_coverage}
\end{figure}

Figure \ref{fig:text_coverage} and Figure \ref{fig:concept_coverage} show the results obtained evaluating  \emph{Text-based Coverage of Wikipedia} and \emph{Concept-based Coverage of Wikipedia}, respectively. These outcomes are useful to measure quality and completeness of the generated summaries with respect to gold summaries. The results have been grouped by tweet streams and for each evaluated approach they are shown with different colors. 

Since \emph{Text-based Coverage of Wikipedia} grows by increasing the level of \varLbl, in Figure \ref{fig:text_coverage} the minimum value produced by TAKE that is higher than the values produced by other approaches has been plotted. 
Specifically, it has been obtained setting the level of \varLabel to $0.6$. For levels of \varLabel greater than $0.6$, TAKE significantly outperforms other approaches revealing good performances in terms of complete description of summarized event at merely syntactically level.

Furthermore, Figure \ref{fig:concept_coverage} shows that TAKE outperforms other techniques in terms of \emph{Concept-based Coverage of Wikipedia}, and so it is possible to conclude that the proposed method reveals good performances  in terms of quality and completeness also at the concept level. 

\begin{figure}[b!]
\centering
\includegraphics[height=6.7in]{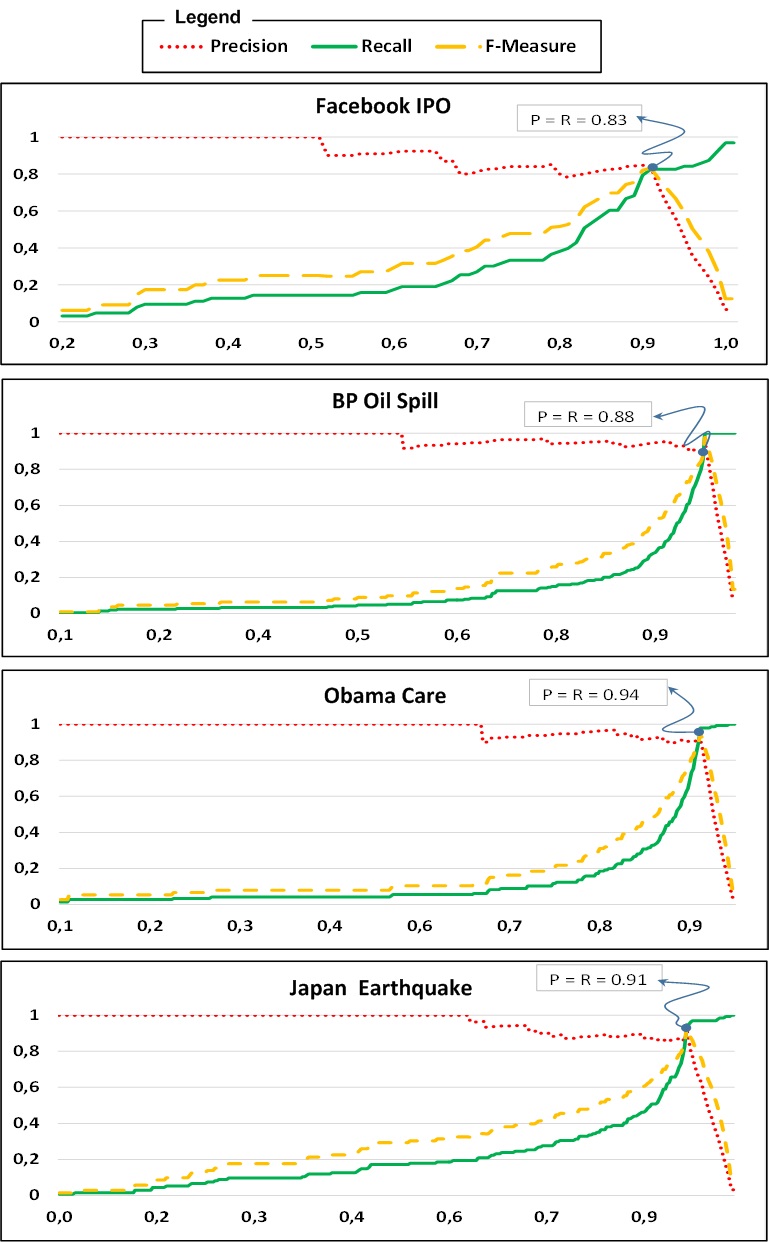}
\caption{Precision, Recall and F-Measure curves of Concept-based Coverage of Wikipedia varying  the level of \varLabel in $[0-1]$.}\label{fig:prcurves}
\end{figure}

In order to provide more details, Figure \ref{fig:prcurves} shows the curves corresponding to Precision, Recall and F-measure of \emph{Concept-based Coverage of Wikipedia} for each tweet stream. It has been evaluated by varying the level of \varLabel \var from $0.0$ to $1.0$. The figure highlights that TAKE reveals valuable performance in terms of Recall with acceptable values of Precision with level of \varLabel between $0.7$ and $0.9$.

In general, the distinguishing feature introduced by TAKE approach is the possibility to have more or less shorten summary ensuring good trade off between quality and completeness both at syntactic and semantic level as shown by the experimental results.

\section{Conclusion}
This work defines Time Aware Knowledge Extraction methodology to support microblog summarization algorithm that has been applied on Twitter. The overall framework relies on Fuzzy Formal Concept Analysis introducing temporal correlation among tweets. Firstly, chronological ordered tweets have been analyzed to detect peaks of microblog activities around a specific topic. Secondly, tweet's content analysis exploits service of wikification 
enabling semantic annotation of the text with wikipedia's entities. Finally, a microblog summarization algorithm has been defined walking across concepts of the resulting timed fuzzy lattice in order to select right tweets covering main concepts of the story and their development over the timeline. Specifically, the distinguishing feature introduced with this work is the level of \varLabel that allows to filter the multitude of the concepts in the timed fuzzy lattice in order to zoom (in or out) the description of specific real world event. The \varLabel enables the users to have more or less verbose update summary according to time constraints.

The framework has been validated comparing the obtained results with other existing methodologies, that are LDA (Latent Dirichlet Allocation), GDTM (Gaussian Decay Topic Model), and DTM (Decay Topic Model). As highlighted in \cite{hp}, these methodologies outperform the LDA baseline by exploiting temporal correlation between tweets and their semantics at two different stages. The proposed framework outperforms the compared approaches considering at the same time temporal correlation among tweets and semantic of their content by means of Time Aware Knowledge Extraction ensuring good trade off between quality, completeness and redundancy. 

Future works can exploit the Time Aware Knowledge Extraction methodology to address challenging research topics in the area of social media analytics, such as topic detection and monitoring, context-aware ad placement, and so on. 
Another interesting future direction is to apply the verification techniques described in \cite{parente} for hierarchical structures to the FCA lattice to verify properties of the concepts.

\section*{Acknowledgement}
The authors are indebted to the authors of \cite{hp} for the provisioning of tweet streams and gold summary in order to enable the comparison evaluation of the proposed framework.

\section*{References}

\bibliography{journal}

\end{document}